\documentclass[twocolumn]{aastex631}

\usepackage{graphicx}
\usepackage{amsmath}

\defcitealias{McLure2018b}{M18}

\shorttitle{KMOS Quiescent Galaxy Stellar Metallicities at $z\gtrsim1$}
\shortauthors{Carnall et al.}

\begin{document}

\title{The stellar metallicities of massive quiescent galaxies at $\mathbf{1.0 < z < 1.3}$ from KMOS+VANDELS}

\correspondingauthor{Adam C. Carnall}
\email{adamc@roe.ac.uk}

\author{Adam C. Carnall}
\affiliation{SUPA, Institute for Astronomy, University of Edinburgh, Royal Observatory, Edinburgh EH9 3HJ, UK}

\author{Ross J. McLure}
\affiliation{SUPA, Institute for Astronomy, University of Edinburgh, Royal Observatory, Edinburgh EH9 3HJ, UK}

\author{James S. Dunlop}
\affiliation{SUPA, Institute for Astronomy, University of Edinburgh, Royal Observatory, Edinburgh EH9 3HJ, UK}

\author{Massissilia Hamadouche}
\affiliation{SUPA, Institute for Astronomy, University of Edinburgh, Royal Observatory, Edinburgh EH9 3HJ, UK}

\author{Fergus Cullen}
\affiliation{SUPA, Institute for Astronomy, University of Edinburgh, Royal Observatory, Edinburgh EH9 3HJ, UK}

\author{Derek J. McLeod}
\affiliation{SUPA, Institute for Astronomy, University of Edinburgh, Royal Observatory, Edinburgh EH9 3HJ, UK}

\author{Ryan Begley}
\affiliation{SUPA, Institute for Astronomy, University of Edinburgh, Royal Observatory, Edinburgh EH9 3HJ, UK}

\author{Ricardo Amorin}
\affiliation{Instituto de Investigaci\'on Multidisciplinar en Ciencia y Tecnolog\'ia, Universidad de La Serena, Ra\'ul Bitr\'an 1305, La Serena, Chile}

\affiliation{Departamento de F\'isica y Astronom\'ia, Universidad de La Serena, Av. Juan Cisternas 1200 Norte, La Serena, Chile}

\author{Micol Bolzonella}
\affiliation{INAF - OAS Bologna, Via P. Gobetti 93/3, I-40129, Bologna, Italy}

\author{Marco Castellano}
\affiliation{INAF - Osservatorio Astronomico di Roma, Via Frascati 33, I-00078 Monteporzio Catone, Italy}

\author{Andrea Cimatti}
\affiliation{University of Bologna, Department of Physics and Astronomy (DIFA), Via Gobetti 93/2, I-40129, Bologna, Italy}

\affiliation{INAF - Osservatorio Astrofisico di Arcetri, Largo E. Fermi 5, I-50125, Firenze, Italy}

\author{Fabio Fontanot}
\affiliation{INAF - Astronomical Observatory of Trieste, via G.B. Tiepolo 11, I-34143 Trieste, Italy}

\affiliation{IFPU - Institute for Fundamental Physics of the Universe, via Beirut 2, 34151, Trieste, Italy}

\author{Adriana Gargiulo}
\affiliation{INAF - IASF Milano, Via A. Corti 12, I-20133, Milano, Italy}

\author{Bianca Garilli}
\affiliation{INAF - IASF Milano, Via A. Corti 12, I-20133, Milano, Italy}

\author{Filippo Mannucci}
\affiliation{INAF - Osservatorio Astrofisico di Arcetri, Largo E. Fermi 5, I-50125, Firenze, Italy}

\author{Laura Pentericci}
\affiliation{INAF - Osservatorio Astronomico di Roma, Via Frascati 33, I-00078 Monteporzio Catone, Italy}

\author{Margherita Talia}
\affiliation{INAF - OAS Bologna, Via P. Gobetti 93/3, I-40129, Bologna, Italy}

\affiliation{University of Bologna, Department of Physics and Astronomy (DIFA), Via Gobetti 93/2, I-40129, Bologna, Italy}

\author{Giovani Zamorani}
\affiliation{INAF - OAS Bologna, Via P. Gobetti 93/3, I-40129, Bologna, Italy}

\author{Antonello Calabro}
\affiliation{INAF - Osservatorio Astronomico di Roma, Via Frascati 33, I-00078 Monteporzio Catone, Italy}

\author{Giovanni Cresci}
\affiliation{INAF - Osservatorio Astrofisico di Arcetri, Largo E. Fermi 5, I-50125, Firenze, Italy}

\author{Nimish P. Hathi}
\affiliation{Space Telescope Science Institute, 3700 San Martin Dr., 
Baltimore, MD 21218, USA}

\begin{abstract}

We present a rest-frame UV-optical ($\lambda=2500{-}6400$\,\AA) stacked spectrum representative of massive quiescent galaxies at $1.0<z<1.3$ with log$(M_*/\mathrm{M_\odot})>10.8$. The stack is constructed using VANDELS survey data, combined with new KMOS observations. We apply two independent full-spectral-fitting approaches, measuring a total metallicity, [Z/H]=$-0.13\pm0.08$ with \textsc{Bagpipes}, and [Z/H]=$0.04\pm0.14$ with \textsc{Alf}, a fall of $\sim0.2-0.3$ dex compared with the local Universe. We also measure an iron abundance, [Fe/H] =$-0.18\pm0.08$, a fall of $\sim0.15$ dex compared with the local Universe. We measure the alpha enhancement via the magnesium abundance, obtaining [Mg/Fe]=$0.23\pm$0.12, consistent with similar-mass galaxies in the local Universe, indicating no evolution in the average alpha enhancement of log$(M_*/\mathrm{M_\odot})\sim11$ quiescent galaxies over the last $\sim8$\,Gyr. This suggests the very high alpha enhancements recently reported for several bright $z\sim1-2$ quiescent galaxies are due to their extreme masses, log$(M_*/\mathrm{M_\odot})\gtrsim11.5$, in accordance with the well-known downsizing trend, rather than being typical of the $z\gtrsim1$ population. The metallicity evolution we observe with redshift (falling [Z/H], [Fe/H], constant [Mg/Fe]) is consistent with recent studies. We recover a mean stellar age of $2.5^{+0.6}_{-0.4}$\,Gyr, corresponding to a formation redshift, $z_\mathrm{form}=2.4^{+0.6}_{-0.3}$. Recent studies have obtained varying average formation redshifts for $z\gtrsim1$ massive quiescent galaxies, and, as these studies report consistent metallicities, we identify different star-formation-history models as the most likely cause. Larger spectroscopic samples from upcoming ground-based instruments will provide precise constraints on ages and metallicities at $z\gtrsim1$. Combining these with precise $z>2$ quiescent-galaxy stellar-mass functions from James Webb Space Telescope will provide an independent test of formation redshifts derived from spectral fitting.

\end{abstract}

\section{Introduction}

In the local, present-day Universe, the massive galaxy population ($\mathrm{log}_{10}(M_*\ /\ \mathrm{M_\odot}) \gtrsim 10.5$) is dominated by quiescent galaxies, which have shut down (quenched) their star-formation activity (e.g. \citealt{McLeod2021}). The formation and quenching processes leading to the rise of this dominant population, across at least the last 12 Gyr since the first known quiescent galaxies at redshift, $z\simeq4$, are therefore of central importance to our understanding of galaxy evolution.

Studying the massive quiescent galaxy population in the local Universe presents several key challenges. Firstly, as the rate of change in stellar population spectra is roughly logarithmic with age (e.g. \citealt{Ocvirk2006}), constraints on ages via spectral fitting become steadily less precise for older stellar populations. This means that formation redshifts for local quiescent galaxies, with stellar population ages $\gtrsim10$ Gyr, are highly uncertain. In addition, massive galaxies gradually accrete new stellar populations via merger events, which have the potential to change their physical sizes, as well as the average ages and metallicities of their stellar populations. This  obscures the signatures of the dominant physical processes that acted on these galaxies during their main epoch of formation.

To understand the rise of massive quiescent galaxies, it is therefore necessary to conduct detailed observational studies across the whole history of the Universe, from the first $1{-}2$ Gyr to the present day. By measuring how the distributions of key physical parameters evolve as a function of redshift, it should be possible to disentangle the degenerate effects of different processes acting at different times, providing strong constraints on the key physical ingredients required to produce the local massive galaxy population.

As a practical first step towards achieving this goal, much attention is currently focused on measuring the number densities of quiescent galaxies, as well as the distributions of their physical sizes, mean stellar ages and metallicities, as a function of stellar mass and redshift. Armed with this information across a large fraction of cosmic history, we may then aspire to build a self-consistent model for the assembly of the quiescent population, including the physical processes that influence massive galaxies during their main epoch of formation and quenching, the subsequent growth of individual quiescent galaxies through merger events, and the growth of the quiescent population as a whole via new galaxies quenching their star-formation activity. 

The largest single factor currently limiting progress in measuring the evolution of these four key properties (number density, size, stellar age and metallicity) across cosmic time is the availability of high-quality observational data for faint, red, massive quiescent galaxies in the high-redshift Universe. The best-constrained parameters for the quiescent population currently are number densities and physical sizes. These are now widely studied out to $z\sim3$, as they can be reliably constrained via high-quality photometric data (e.g. \citealt{McLure2013}; \citealt{vanderwel2014}; \citealt{Straatman2014, Straatman2016}; \citealt{Cecchi2019}; \citealt{Mowla2019a, Mowla2019b}; \citealt{Suess2019a, Suess2019b}; \citealt{Girelli2019}; \citealt{Merlin2019}; \citealt{Sherman2020}; \citealt{Carnall2020}; \citealt{Marsan2020}; \citealt{Santini2021}; \citealt{Hamadouche2022}). 

Stellar ages (more generally star-formation histories; SFHs) have also been studied using photometric data (e.g. \citealt{Pacifici2016}; \citealt{Carnall2018}). However, the age-metallicity-dust degeneracy in galaxy spectral shapes results in relatively weak constraints, meaning the applied priors significantly impact the results obtained (\citealt{Carnall2019a}; \citealt{Leja2019a}). Recently, the increasing availability of medium to high signal-to-noise ratio (SNR) spectroscopic data for quiescent galaxies at intermediate redshifts (e.g. \citealt{vanderWel2016}; \citealt{McLure2018b} - hereafter \citetalias{McLure2018b}), combined with sophisticated full-spectral-fitting approaches, have produced the first strong constraints on the SFHs of representative samples out to $z\gtrsim1$ (e.g. \citealt{Wu2018a, Wu2018b}; \citealt{Belli2019}; \citealt{Carnall2019b}; \citealt{Estrada-Carpenter2019, Estrada-Carpenter2020}; \citealt{Wild2020}; \citealt{Tacchella2021}). 

However, systematic differences still exist in the results of these studies, likely due to a combination of different assumed SFH models, and the fact that stellar metallicities (which are strongly degenerate with ages) are relatively weakly constrained by intermediate-SNR data, meaning the applied priors still play a significant role. 

Measuring the stellar metallicities of quiescent galaxies represents another step-change in observational difficulty, with strong constraints ($\sim0.1$ dex) only available at continuum SNR $\gtrsim10{-}15\,$\AA$^{-1}$ in the rest-frame optical (e.g. \citealt{Gallazzi2005}; \citealt{Pacifici2012}). This is compounded at $z \gtrsim 1$ by the key features being shifted into the near-IR, where stronger atmospheric absorption and emission make continuum observations from the ground far more challenging.

Despite these challenges, stellar metallicity measurements are highly valuable, as they are strongly constraining on galaxy-formation models, being intimately linked to the physics of star formation and gas recycling (e.g. \citealt{Maiolino2019}). Whilst substantial progress has been made out to $z\sim3$ with the brighter, bluer continua of star-forming galaxies (e.g. \citealt{Steidel2016}; \citealt{Cullen2019, Cullen2021}), to date, studies of quiescent-galaxy stellar metallicities at $z>1$ are rare, and often restricted to either individual bright objects, or low-resolution grism spectra (e.g. \citealt{Whitaker2013}; \citealt{Onodera2015}; \citealt{Lonoce2015}; \citealt{Kriek2016}). 

Whilst the upcoming \textit{James Webb Space Telescope} (JWST) will produce exceptionally high-quality data for limited numbers of objects at $z>1$, truly statistical studies are still several years away, awaiting the advent of the next generation of ground-based multi-object spectrographs (e.g. \citealt{Cirasuolo2020}).

In this work, we present the first determination of the stellar metallicities of a mass-selected sample of quiescent galaxies at $z > 1$. We combine rest-frame near-UV data from the VANDELS survey (\citetalias{McLure2018b}; \citealt{Pentericci2018}; \citealt{Garilli2021}) with new rest-frame optical KMOS $YJ-$band spectroscopy, to produce a stacked spectrum covering rest-frame $2500{-}6400$\,\AA\ for UVJ-selected galaxies at $1.0 < z < 1.3$ with \hbox{$\mathrm{log_{10}(}M_*/\mathrm{M_\odot)} > 10.8$}. We fit our stack using both \textsc{Bagpipes} \citep{Carnall2018} and \textsc{Alf} \citep{Conroy2012, Conroy2018}, obtaining consistent ages and metallicities.

At $1.0 < z < 1.3$, the VANDELS spectra span rest-frame wavelengths, $\lambda\sim2500{-}4500$\,\AA. This means they include the Balmer/4000\,\AA\ break region critical for precise age determination. However, generally, they do not include the key Fe and Mg absorption features at $4500{-}5500$\,\AA\ most commonly used to measure stellar metallicities in the local Universe (e.g. \citealt{Gallazzi2005}).

Stellar metallicity was fitted as a free parameter in the VANDELS full-spectral-fitting analysis of \cite{Carnall2019b}. However, due to the lack of these key Fe and Mg features, combined with the fact that empirical stellar-population models, generally still accepted as more accurate than those based on theoretical spectra (e.g. \citealt{Coelho2020}), are only available at $\gtrsim3500$\,\AA, we chose not to report our derived stellar metallicities in that work. Instead, we obtained further observations with KMOS to gain access to these key rest-frame optical features, to ensure reliable results, and to facilitate direct comparisons with previous studies in the local Universe.

The structure of this paper is as follows. In Sections \ref{vandels} and \ref{kmos}, we introduce our VANDELS and KMOS spectroscopic datasets respectively. In Section \ref{stack}, we describe the selection of our mass-complete sample, as well as the process of constructing and fitting our representative stacked spectrum with \textsc{Bagpipes} and \textsc{Alf}. Our results are presented in Section \ref{results}, and discussed in Section \ref{discussion}. We present our conclusions in Section \ref{conclusion}. All magnitudes are quoted in the AB system. For cosmological calculations, we adopt $\Omega_M = 0.3$, $\Omega_\Lambda = 0.7$ and $H_0$ = 70 $\mathrm{km\ s^{-1}\ Mpc^{-1}}$. All times, $t$, are measured forwards from the beginning of the Universe. We assume a \cite{Kroupa2001} initial mass function. We also assume the Solar abundances of \cite{Asplund2009}, such that $\mathrm{Z_\odot} = 0.0142$.

\section{VANDELS data and sample selection}\label{vandels}

VANDELS (\citetalias{McLure2018b}; \citealt{Pentericci2018}; \citealt{Garilli2021}) is a large ESO public spectroscopic survey of the high-redshift Universe, using the VIsible Multi Object Spectrograph (VIMOS) instrument on the Very Large Telescope (VLT). The primary aim of VANDELS is to detect continuum emission at high SNR for high-redshift galaxies, moving beyond redshift acquisition to study galaxy physical properties within the first 6 billion years prior to $z=1$.

\subsection{The parent photometric sample}\label{vandels:parent}

The parent photometric sample for this study consists of 812 massive quiescent galaxies, selected by \citetalias{McLure2018b} as potential targets for the VANDELS survey. These objects were selected from four photometric catalogues: the CANDELS GOODS South and UDS catalogues of \cite{Guo2013} and \cite{Galametz2013}, and two further ground-based photometric catalogues, purpose-built for VANDELS. These cover the regions immediately surrounding the CANDELS footprints. This approach was necessary as the VIMOS spectrograph field of view covers a larger area than the CANDELS HST imaging.

The parent sample was selected from these four photometric catalogues by the following process (described in full detail in section 4 of \citetalias{McLure2018b}). Objects were first required to meet the following apparent magnitude and photometric redshift, $z_\mathrm{phot}$, criteria
\begin{itemize}
    \setlength\itemsep{0.5em}
    \item $H \leq 22.5$
    \item $i \leq 25$
    \item $1.0 \leq z_\mathrm{phot} \leq 2.5$.
\end{itemize}

\noindent For the CANDELS catalogues, the photometric redshifts used were those published by the CANDELS team \citep{Dahlen2013}. For the ground-based catalogues, photometric redshifts were generated through a similar process by the VANDELS team, taking the median of results obtained using a variety of public codes. The $H$-band magnitude cut limits the sample to objects with \hbox{$\mathrm{log_{10}(}M_*/\mathrm{M_\odot)} \gtrsim 10$}. The $i$-band cut was implemented to make sure the faintest objects would be detected in the VANDELS spectra, and is not relevant to this study, which is focused on a brighter, mass-selected sub-sample.

To select only quiescent galaxies, the following, permissive, rest-frame UVJ magnitude selection criteria were then applied:
\begin{itemize}
    \setlength\itemsep{0.5em}
    \item $U-V > 0.88(V-J) + 0.49$
    \item $V-J < 1.6$
    \item $U-V > 1.2$
\end{itemize}

\noindent From the 812 galaxies selected, 64 per cent have $z_\mathrm{phot} < 1.3$, with the sample being mass complete down to $\mathrm{log_{10}(}M_*/\mathrm{M_\odot)} = 10.3$ at $1.0 < z < 1.3$ \citep{Carnall2019b}. These 812 objects are referred to as the parent sample throughout the rest of this work.

\subsection{VANDELS spectroscopic observations}\label{vandels:spec}

From the parent sample described in the previous sub-section, objects were assigned to slits at random to be observed as part of VANDELS. Spectra were obtained for 281 massive quiescent galaxies, roughly one third of the parent sample. These represent roughly 13 per cent of the VANDELS survey, which includes $\sim2000$ objects in total. The remaining 87 per cent is composed of star-forming galaxies at $z > 2.4$. Objects were observed for 20, 40 or 80 hours, depending on their $i$-band magnitudes, with a mean integration of $\sim45$ hours.

The VANDELS spectroscopic data (described in full in \citealt{Pentericci2018}) covers observed-frame wavelengths from $4800{-}10300$\,\AA, with spectral resolving power, $R=\lambda/\Delta\lambda\sim600$. The VANDELS data have high SNR for these redshifts. The spectra of quiescent galaxies within the mass-complete section of the parent sample discussed in Section \ref{vandels:parent} (those with \hbox{$1.0 < z < 1.3$} and \hbox{$\mathrm{log_{10}(}M_*/\mathrm{M_\odot)} > 10.3$}) have a median continuum SNR $\sim$ 11\,\AA$^{-1}$ at 7500\,\AA. 

Spectroscopic redshifts were measured by the VANDELS team using the pandora.ez software \citep{Garilli2010}. Spectra were assigned redshift quality flags following \cite{LeFevre2013}, with all but 12 of the 281 observed massive quiescent galaxies assigned quality flags 3 or 4, corresponding to 95 and 100 per cent probabilities of correct identification respectively (in fact, \citealt{Garilli2021} demonstrate $\sim99$ per cent reliability for the VANDELS flag 3 and 4 redshifts combined). The 12 spectra with lower quality flags, as well as one low-redshift interloper, were excluded from this work, leaving 268 spectra. Three objects with spectroscopic redshifts in the range $0.96 < z < 1.0$ were retained. All VANDELS spectroscopic data used in this work comes from the final public data release, DR4, described in \cite{Garilli2021}.

\section{KMOS Data and Sample Selection}\label{kmos}

We observed four KMOS pointings during ESO P104, from October 2019\,$-$\,January 2020, under programme ID 0104.B$-$0885(A). Two pointings targeted each of the two VANDELS fields (UDS and GOODS South). Observations were in the $YJ-$band, providing wavelength coverage from $\sim1-1.35\,\mathrm{\mu m}$ at $R\sim3600$. During the period in which our data were taken, 23 KMOS arms were functional, meaning we obtained spectra for a total of 92 objects. Each pointing was observed for 8 hours on source, with 300-second exposures, and an ABAB nodding pattern between object and sky positions. The mean seeing for our observations was $0.6^{\prime\prime}$.

\subsection{KMOS sample selection}\label{kmos:sample}

Targets for our KMOS observations were drawn almost exclusively from the parent sample of 812 objects described in Section \ref{vandels:parent}. To maximise the utility of our $YJ-$band observations, objects were prioritised by their $J-$band magnitude. Objects were assigned to KMOS IFUs using the KMOS ArM Allocator (KARMA) tool, with priority classes being defined as follows

\begin{enumerate}
\item Parent sample with $J < 21.5$
\item Parent sample with $J > 21.5$
\item Post-starburst galaxies from \cite{Wilkinson2021}.
\end{enumerate}

\noindent The priority 3 targets, obtained via private communication, are unrelated to this study, and were added only as fillers to make sure all IFUs were assigned. The positions of the four KMOS pointings were optimised to target the maximum possible number of priority class 1 objects, as well as to maximise the overlap with VANDELS spectroscopy. In total, we allocated IFUs to 62 objects with priority 1, 29 objects with priority 2, and 1 object with priority 3.

From the parent sample of 812 objects, a total of 273 have \hbox{$J < 21.5$}, meaning objects of priority class 1 were $\sim 4$ times more likely to be observed than those of priority class 2. From the 91 parent sample objects observed with KMOS, 51 objects were also observed by VANDELS, as described in Section \ref{vandels}.

\subsection{KMOS data reduction}

Our KMOS data were reduced using a combination of the standard \textsc{Esorex} pipeline recipes and custom code, optimised for these data. The KMOS pipeline v4.0.0 was used to produce flux-calibrated, sky-subtracted cubes for each individual pair of 300 second (object, sky) exposures. All our targets with $J \lesssim 22$ (approximately 75 out of 92 objects, including all those used in this paper) are clearly detected in each single-exposure, wavelength-collapsed cube.

Upon inspection of the wavelength-collapsed cubes, $1{-}2$ spaxel ($0.2{-}0.4^{\prime\prime}$) shifts in object centroids were noted with respect to the WCS coordinates of each object. New centroid positions were therefore measured for each frame by selecting the brightest pixel within a 1$^{\prime\prime}-$diameter circular aperture centred on the WCS position. These new centroids were verified by manual inspection of each cube.

Also noted were substantial sky-line residuals still present in the data cubes, along with varying systematic shifts in pixel values away from zero across the whole wavelength axis. To address these issues, we implement a further, custom, sky-subtraction step.  We first mask all pixels within a 1$^{\prime\prime}-$diameter circular aperture centred on the new object centroid pixel, as well as all pixels bordering the edge of the detector. We then subtract the median of the remaining pixels from the cube slice at each wavelength. This approach is only possible in this specific instance, as all our targets, being $z>1$ quiescent galaxies, are extremely compact, with effective radii, $r_\mathrm{e} << 1^{\prime\prime}$ (e.g. \citealt{McLure2013}; \citealt{vanderwel2014}).

Exposures were then aligned according to their updated centroid positions and median stacked, with uncertainties calculated via the robust median absolute deviation (MAD) indicator. Finally, 1D spectra were extracted within 1$^{\prime\prime}-$diameter circular apertures, using the \cite{Horne1986} optimal extraction algorithm.

Both the custom re-centroiding and additional sky-subtraction steps are critical to recovering the expected Fe, Mg and Na absorption features in the spectra of individual objects. Using the default, pipeline-combined cubes only produces visible continuum features in the spectrum of our brightest target, with $J=19.6$. However, with these additional steps, the strongest features (Mg\,\textsc{i} 5170\,\AA, and Na\,\textsc{i} 5895\,\AA) are visible in almost all spectra for objects with $J < 21.5$.

\subsection{KMOS redshift measurement}\label{kmos:redshift}

As described above, 51 out of 91 objects from the parent sample observed with KMOS already have secure spectroscopic redshifts from VANDELS. For the remaining 40 objects, we followed the same process as described in Section \ref{vandels:spec}, using Pandora.ez, to measure redshifts. In Table \ref{table:redshifts}, we report 25 new spectroscopic redshifts, $z_\mathrm{KMOS}$, measured from our KMOS data, along with their associated quality flags. The remaining 15 objects are all significantly fainter ($J > 21.8$), meaning that no reliable redshifts could be measured.

\begin{figure}
	\includegraphics[width=\columnwidth]{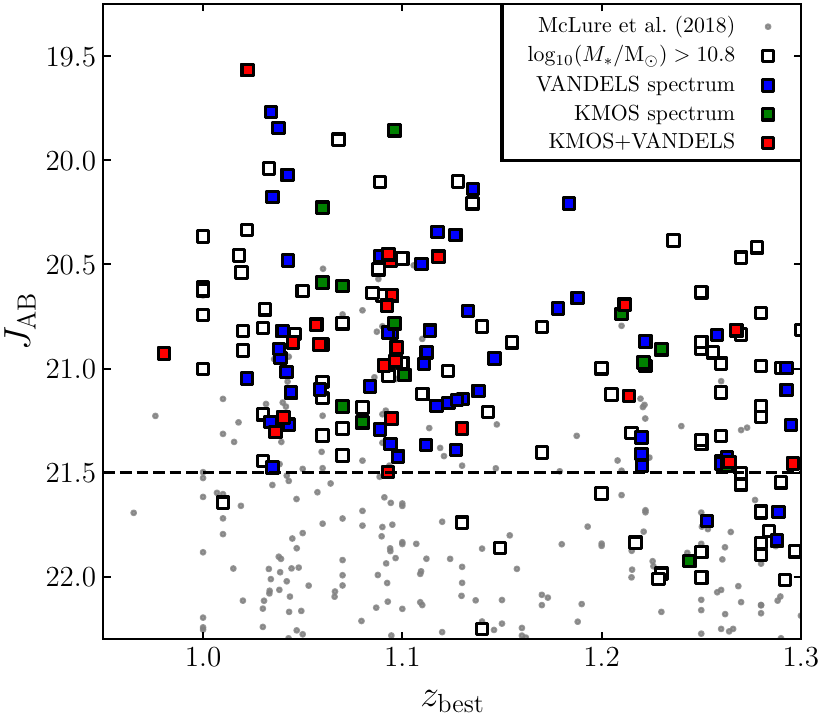}
    \caption{The distribution of the sample in $J-$band magnitude versus redshift. Circles denote the parent sample from \protect\citetalias{McLure2018b} (see Section \ref{vandels:parent}). Objects included in our mass-complete photometric sample, selected as described in Section \ref{stack:sel}, are marked with squares. The colours denote the availability of spectroscopic data from VANDELS and KMOS for each object. The dashed horizontal line is the threshold for inclusion as a priority 1 target in our KMOS observations (see Section \ref{kmos:sample}). Objects below were assigned priority 2.}
    \label{fig:mag_vs_z}
\end{figure}

\begin{table}
\movetableright=-0.5in
  \caption{Spectroscopic redshifts measured from our KMOS data (see \hbox{Section \ref{kmos:redshift})}. From the 91 parent sample objects observed with KMOS, 51 already have robust VANDELS spectroscopic redshifts. The 25 objects in this table are those of the remaining 40 for which redshifts could be measured. The remaining 15 objects are all fainter than $J = 21.8$.
}\label{table:redshifts}
\begingroup
\setlength{\tabcolsep}{4pt} 
\begin{tabular}{lccccc}
\hline
ID & RA & DEC & $J$ & $z_\mathrm{KMOS}$ & Flag\\
\hline
CDFS-017418&53.15497&$-$27.76891&19.86&1.0956&4 \\
UDS-196414&34.48731&\phantom{0}$-$5.09687&20.23&1.0921&4 \\
UDS-205452&34.50079&\phantom{0}$-$5.05545&20.59&1.0945&4 \\
UDS-190420&34.51167&\phantom{0}$-$5.12379&20.60&1.0945&3 \\
UDS-200825&34.52746&\phantom{0}$-$5.07677&20.74&1.0369&3 \\
UDS-201280&34.51318&\phantom{0}$-$5.07575&20.78&1.4140&2 \\
CDFS-004529&53.08042&$-$27.87204&20.78&1.0967&3 \\
UDS-196179&34.50759&\phantom{0}$-$5.09889&20.91&1.2710&3 \\
CDFS-016336&53.22897&$-$27.77253&20.95&1.0386&3 \\
CDFS-020067&53.15878&$-$27.74239&20.97&1.2210&3 \\
CDFS-022694&53.14845&$-$27.71946&20.99&1.2221&3 \\
CDFS-003952&53.07276&$-$27.87632&21.03&1.1014&4 \\
CDFS-004376&53.07153&$-$27.87246&21.05&1.0976&3 \\
UDS-197616&34.43864&\phantom{0}$-$5.09219&21.07&1.6523&4 \\
UDS-015126&34.31596&\phantom{0}$-$5.19366&21.09&1.3611&2 \\
UDS-207822&34.48340&\phantom{0}$-$5.04458&21.18&1.0334&3 \\
UDS-024934&34.52885&\phantom{0}$-$5.12719&21.22&1.0954&3 \\
UDS-009642&34.33053&\phantom{0}$-$5.22374&21.26&1.0735&3 \\
UDS-013785&34.32431&\phantom{0}$-$5.20136&21.27&1.0915&4 \\
UDS-013519&34.33534&\phantom{0}$-$5.20168&21.40&1.5321&3 \\
UDS-192952&34.48285&\phantom{0}$-$5.11423&21.43&1.3295&3 \\
UDS-020224&34.34611&\phantom{0}$-$5.16692&21.44&1.0830&2 \\
UDS-005970&34.31475&\phantom{0}$-$5.24326&21.45&1.7127&2 \\
UDS-010643&34.38065&\phantom{0}$-$5.21789&21.46&1.2624&3 \\
CDFS-014839&53.16516&$-$27.78587&21.53&1.3175&2 \\

\hline
\end{tabular}
\endgroup
\end{table}

\begin{figure*}
	\includegraphics[width=\textwidth]{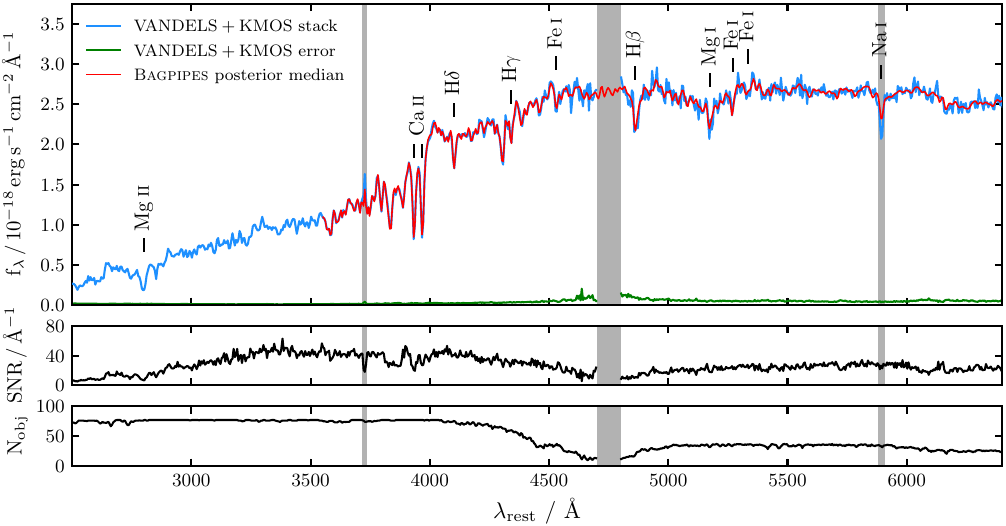}
	
    \caption{The stacked rest-frame UV-optical spectrum for quiescent galaxies at $1.0 < z < 1.3$ with $\mathrm{log_{10}(}M_*/\mathrm{M_\odot)} > 10.8$. The stacked spectrum is shown in blue in the top panel, with the error spectrum shown in green. The SNR per \AA\ and number of objects contributing to the stack are shown in the middle and lower panels respectively. Key age and metallicity sensitive features are marked in black. The red line overlaid on the spectrum shows the posterior median model fitted to the stack with \textsc{Bagpipes}, as described in Section \ref{stack:fit_bagpipes}. This includes a calibration polynomial, which makes corrections of $\sim 5$ per cent. We derive a stellar metallicity of [Z/H] = $-0.13\pm0.08$ and a formation redshift of $z_\mathrm{form} = 2.4^{+0.6}_{-0.3}$. The gray vertical bands show regions masked from fitting (see Section \ref{stack}).}
    \label{fig:stack}
\end{figure*}

\section{Stacking analysis}\label{stack}

The KMOS spectra we obtained for objects with $J < 21.5$ have a median \hbox{SNR $\sim4$\,\AA$^{-1}$} at 12000\,\AA\ ($\sim14$ per resolution element at $R=1000$). This is unfortunately not sufficient to constrain the stellar metallicities of individual objects. In this section, we therefore define a mass-complete sample, for which we can construct a representative stacked spectrum from the VANDELS + KMOS data described in Sections \ref{vandels} and \ref{kmos}. We fit these data to constrain the average stellar metallicities of massive quiescent galaxies at $1.0 < z< 1.3$.

\subsection{Sample selection}\label{stack:sel}

We begin by re-fitting the photometric data described in Section \ref{vandels:parent} for the 812 objects in the \citetalias{McLure2018b} parent sample with \textsc{Bagpipes}. We use the best available redshifts, $z_\mathrm{best}$, defined by the following ranking:

\begin{enumerate}
\item VANDELS DR4 redshift if flag 3 or 4; else
\item KMOS redshift (Table \ref{table:redshifts}) if flag 3 or 4; else
\item Other spectroscopic redshift (see \citetalias{McLure2018b}); else
\item Photometric redshift, described in Section \ref{vandels:parent}.
\end{enumerate}

\noindent For consistency with other published work, we assume a \cite{Calzetti2000} dust attenuation curve and Solar metallicity. We fit for five free parameters: $V-$band attenuation ($A_V$), total stellar mass formed, and the three shape parameters of a double-power-law SFH model (e.g. \citealt{Carnall2018}). The priors assumed for these five parameters are the same as shown in Table \ref{table:params}, which provides the full list of parameters and priors we use for our more-sophisticated fits to our stacked spectrum plus photometry in Section \ref{stack:fit_bagpipes}.

\textsc{Bagpipes} uses the 2016 updated version of the \cite{Bruzual2003} stellar population models\footnote{\url{https://www.bruzual.org/~gbruzual/bc03/Updated_version_2016}}, using the MILES stellar spectral library (\citealt{Falcon-Barroso2011}) and updated stellar evolutionary tracks of \cite{Bressan2012} and \cite{Marigo2013}. 

From these fits, we obtain stellar masses, $M_*$, and rest-frame UVJ magnitudes. We now define a mass-complete sample for which we can construct a representative stacked spectrum from our combined VANDELS + KMOS spectroscopic datasets. We begin by imposing $z_\mathrm{best} < 1.3$, leaving 512 objects. We then follow up on the permissive UVJ criteria of \citetalias{McLure2018b} (see Section \ref{vandels:parent}) by requiring \hbox{$(U - V) > 0.88\times(V - J) + 0.69$}. This is the diagonal UVJ cut proposed by \cite{Williams2009} for quiescent galaxy selection at $z < 0.5$. We use this criterion for our $z\gtrsim1$ sample, as it has been shown by \cite{Carnall2018, Carnall2019b} to consistently select objects with sSFR $< 0.2$ $t_\mathrm{H}^{-1}$ across a wide redshift range (where $t_\mathrm{H}$ is the age of the Universe as a function of redshift). This is a widely used criterion for separating star-forming and quiescent galaxies (e.g. \citealt{Pacifici2016}). This further reduces the sample to 409 objects.

As our KMOS observations targeted a more-limited, brighter sub-sample than VANDELS, the selection criteria detailed in Section \ref{kmos:sample} are the most important for defining our mass-completeness limit. We define this limit as the lowest stellar mass for which 90 per cent of more massive galaxies have $J < 21.5$, which is the criterion for inclusion as a priority 1 target in our KMOS observations. This lowest mass is approximately \hbox{$\mathrm{log_{10}(}M_*/\mathrm{M_\odot)} = 10.8$}, and we therefore adopt this as our mass-completeness limit. Imposing this stellar-mass criterion returns 176 objects, with a median stellar mass of \hbox{$\mathrm{log_{10}(}M_*/\mathrm{M_\odot)} = 11.05$}. VANDELS spectroscopy is available for 77 of these, whereas KMOS spectroscopy is available for 37 objects. A total of 23 objects have both VANDELS and KMOS spectra. In both cases, objects with spectroscopic data represent a random draw from our 176-object mass-complete sample. All VANDELS and KMOS objects have secure (flag 3 or 4) spectroscopic redshifts.

Fig. \ref{fig:mag_vs_z} shows the distribution of galaxies in $J-$band magnitude versus $z_\mathrm{best}$. The \citetalias{McLure2018b} parent sample is shown with gray circles, whereas objects in our mass-complete sample are highlighted with open black squares. The availability of VANDELS and KMOS spectroscopic data is indicated by different coloured fills of these open black squares, as indicated in the figure.

To summarise, our mass-complete sample is selected from the \citetalias{McLure2018b} parent sample, introduced in Section \ref{vandels:parent}, as follows

\begin{itemize}

\item $1.0 < z_\mathrm{best} < 1.3$
\item $\mathrm{log_{10}(}M_*/\mathrm{M_\odot)} > 10.8$
\item $(U - V) > 0.88\times(V - J) + 0.69$

\end{itemize}

\noindent with $M_*$ and UVJ magnitudes determined by \textsc{Bagpipes} fitting.

\subsection{Stacking procedure}

We stack the 77 VANDELS and 37 KMOS spectra for our mass-complete sample to produce a single representative stacked spectrum for massive quiescent galaxies at $1.0 < z < 1.3$ with \hbox{$\mathrm{log_{10}(}M_*/\mathrm{M_\odot)} > 10.8$}. The individual VANDELS and KMOS spectra were first shifted to the rest-frame, then flux-normalised, using rest-frame wavelengths, $\lambda=2500{-}4750$\,\AA\ for VANDELS spectra and $\lambda=5000{-}6000$\,\AA\ for KMOS spectra.

The spectra were then resampled to a common wavelength grid using \textsc{SpectRes} (\citealt{Carnall2017}). The VANDELS and KMOS spectra were then median stacked separately to produce two stacks, with uncertainties calculated via the MAD estimator. Pixels with strong sky line contamination were masked prior to stacking. The two stacked spectra were then multiplied by the median of the normalisation factors applied to their input spectra. The two stacks, which do not overlap in wavelength, were then combined, and finally binned down to 5\,\AA\ sampling.

The stack covers rest-frame wavelengths from $2500{-}6400$\,\AA, with the transition from VANDELS to KMOS data at $\simeq4750$\,\AA. Fewer than 10 objects have wavelength coverage between $4700{-}4800$\,\AA, and we therefore mask this region from the stack. We do not attempt to match the flux normalisations of the two stacks at this stage. Instead, during the fitting procedure described in Section \ref{stack:fit_bagpipes}, we fit a spectrophotometric calibration polynomial to both sections of the stack separately, allowing their relative normalisations to be fitted.

The combined stacked spectrum is shown in blue in the top panel of Fig.~\ref{fig:stack}. The error spectrum is shown in green in the same panel. The SNR per \AA\ and the number of objects contributing to the stack are shown in the lower two panels. The wavelengths of key age and metallicity sensitive absorption features are labelled in black.

We also generate stacked photometry for our mass-complete sample, by taking the posterior median model fitted to each of the 176 objects in Section \ref{stack:sel}, shifting this to the median redshift of our sample ($z = 1.15$), then calculating fluxes through a series of UV-IR filters. We use $UBVRiz$, HST F606W, F814W, F125W and F160W, the HAWKI $K-$band, and $Spitzer-$IRAC Channels 1 and 2. We then produce stacked photometry following the same normalisation and median stacking process as detailed above for the spectroscopic data.

\subsection{BAGPIPES fitting of the stacked spectrum and photometry}\label{stack:fit_bagpipes}

To constrain the average stellar metallicity of our mass-complete sample, we fit our stacked spectrum and photometry with \textsc{Bagpipes}\footnote{\url{https://bagpipes.readthedocs.io}} (\citealt{Carnall2018}). We apply the fitting methodology developed in \cite{Carnall2019b}, described in full detail in section 4 of that work. We here provide a brief summary of the method, including a description of the minor changes that have been made to the fitted model for this work. A full list of the 19 free parameters of our model, along with their associated priors, is given in Table \ref{table:params}.

We fit a double-power-law SFH, this time allowing stellar metallicity to vary with a logarithmic prior from $0.01{-}2.5$\,Z$_\odot$.  Dust attenuation is modelled using the form of \cite{Salim2018}, which parameterises dust curve shape with a power-law deviation, $\delta$, from the \cite{Calzetti2000} model.  Emission lines are included in the fit, using a method based on that of \cite{Byler2017} with the \textsc{Cloudy} photoionization code (\citealt{Ferland2017}). The lifetime assumed for the stellar birth clouds giving rise to nebular emission is 10 Myr, and $A_V$ is doubled for emission from stars younger than this, as well as nebular line and continuum emission.

Two separate second-order multiplicative Chebyshev polynomials are fitted to the VANDELS and KMOS portions of the stacked spectrum to model any imperfections in spectrophotometric calibration. We model the covariance matrix for our spectroscopic data as follows. The diagonal terms are given by the square of the green error spectrum plotted in Fig. \ref{fig:stack}, multiplied by a factor, $a^2$, to allow for potential underestimation of uncertainties. We fit $a$ with a logarithmic prior from $0.1{-}10$. The off-diagonal terms are modelled with a Gaussian process, using an exponential-squared kernel.

The stacked spectrum is shifted to the median redshift of our sample ($z=1.15$) for fitting, and redshift is allowed to vary within a narrow range about this value. A Gaussian prior is applied, with a standard deviation of 0.001, and a maximum deviation of 0.005.

When performing full spectral fitting on spectroscopic observations of galaxies, it is critical to model broadening of spectral features as a result of stellar velocity dispersion within the galaxy, as well as instrumental broadening due to the finite spectral resolution of the optical system. In \textsc{Bagpipes}, these effects are treated as a nuisance parameter, and jointly modelled by convolving the spectral model with a Gaussian kernel in velocity space, with the standard deviation allowed to vary.

Because the VANDELS and KMOS spectra have different spectral resolution, it was initially unclear whether it would be appropriate to fit the whole stack using a single Gaussian kernel, or whether separate kernels for both sections of the stack would be more appropriate. To investigate this, we performed separate fits to the two sections of the stack, obtaining consistent values for the standard deviation. We therefore proceeded with our final fit using a single Gaussian kernel, permitting the standard deviation to vary from $100{-}500$ km\ s$^{-1}$ with a logarithmic prior. The resulting values are not used in our analysis, and we make no attempt to correct these values for instrumental effects to obtain the true stellar velocity dispersion.

We exclude rest-frame wavelengths $<3550$\,\AA\ from the fit, as the MILES library does not provide coverage bluer than this, with the \cite{Bruzual2003} models instead employing a combination of theoretical stellar spectral libraries. We mask the [O\,\textsc{ii}] line at 3727\,\AA, as it is currently unclear whether [O\,\textsc{ii}] emission in quiescent galaxies originates from star-forming regions, which is the only source of line emission in our \textsc{Bagpipes} model. Finally, we also mask the Na D absorption feature at 5895\,\AA, as this has a potential strong interstellar medium component (however, see \citealt{Conroy2014}).

To sample the posterior distribution for a model, \textsc{Bagpipes} uses the \textsc{MultiNest} nested sampling algorithm (\citealt{Skilling2006}; \citealt{Feroz2008, Feroz2009, Feroz2019}), via the \textsc{PyMultiNest} interface (\citealt{Buchner2014}). For fitting our stacked data, \textsc{MultiNest} was run with 1000 live points, requiring $\sim500$ CPU hours.

\begin{table*}
\movetableright=-0.7in
  \caption{Parameters and priors for the \textsc{Bagpipes} model we fit to our stacked spectrum and photometry. The model is described in Section \ref{stack:fit_bagpipes}, and is adapted from the model presented in section 4 of \protect\cite{Carnall2019a}. The first 10 parameters describe our physical model, whereas the final 9 describe our systematic uncertainties model. For Gaussian priors, $\mu$ is the mean and $\sigma$ the standard deviation. Logarithmic priors are uniform in log base ten of the parameter.}
\begingroup
\setlength{\tabcolsep}{4pt}
\renewcommand{\arraystretch}{1.05}
\begin{tabular}{lllllll}
\hline
Component & Parameter & Symbol / Unit & Range & Prior & \multicolumn{2}{l}{Hyperparameters} \\
\hline
Global & Redshift & $z$ & (1.145, 1.155) & Gaussian &  $\mu = 1.15$ & $\sigma$ = 0.001 \\
  & Velocity dispersion/instrumental broadening & $\sigma_\mathrm{vel}$ / km s$^{-1}$ & (100, 500) & logarithmic & & \\
\hline
SFH & Stellar mass formed & $M_*\ /\ \mathrm{M_\odot}$ & (1, $10^{13}$) &logarithmic & & \\
 & Metallicity & $Z\ /\ \mathrm{Z_\odot}$ & (0.01, 2.5) &logarithmic & & \\
 & Falling slope & $\alpha$ & (0.1, 1000) & logarithmic & & \\
 & Rising slope & $\beta$ & (0.1, 1000) & logarithmic & & \\
 & Peak time & $\tau$ / Gyr & (0.1, $t_\mathrm{H}$) & uniform & & \\
 \hline
Dust & $V-$band attenuation & $A_V$ / mag & (0, 4) & uniform & & \\
& Deviation from \cite{Calzetti2000} slope & $\delta$ & ($-0.3$, 0.3) & Gaussian & $\mu = 0$ & $\sigma$ = 0.1 \\
& Strength of 2175\,\AA\ bump & $B$ & (0, 5) & uniform & & \\
\hline
Calibration & VANDELS zero order & $P_{V0}$ & (0.5, 1.5) & Gaussian & $\mu = 1$ & $\sigma$ = 0.25 \\
& VANDELS first order & $P_{V1}$ & ($-0.5$, 0.5) & Gaussian & $\mu = 0$ & $\sigma$ = 0.25\\
& VANDELS second order & $P_{V2}$ & ($-0.5$, 0.5) & Gaussian & $\mu = 0$ & $\sigma$ = 0.25\\
& KMOS zero order & $P_{K0}$ & (0.5, 1.5) & Gaussian & $\mu = 1$ & $\sigma$ = 0.25 \\
& KMOS first order & $P_{K1}$ & ($-0.5$, 0.5) & Gaussian & $\mu = 0$ & $\sigma$ = 0.25\\
& KMOS second order & $P_{K2}$ & ($-0.5$, 0.5) & Gaussian & $\mu = 0$ & $\sigma$ = 0.25\\
\hline
Noise & White noise scaling & $a$ & (0.1, 10)  & logarithmic & & \\
 & Correlated noise amplitude & $b$ / $f_\mathrm{max}$ & (0.0001, 1)  & logarithmic & & \\
 & Correlation length & $l$ / $\Delta\lambda$  & (0.01, 1)  & logarithmic & & \\
\hline
\end{tabular}
\endgroup
\label{table:params}
\end{table*}

\begin{figure*}
	\includegraphics[width=\textwidth]{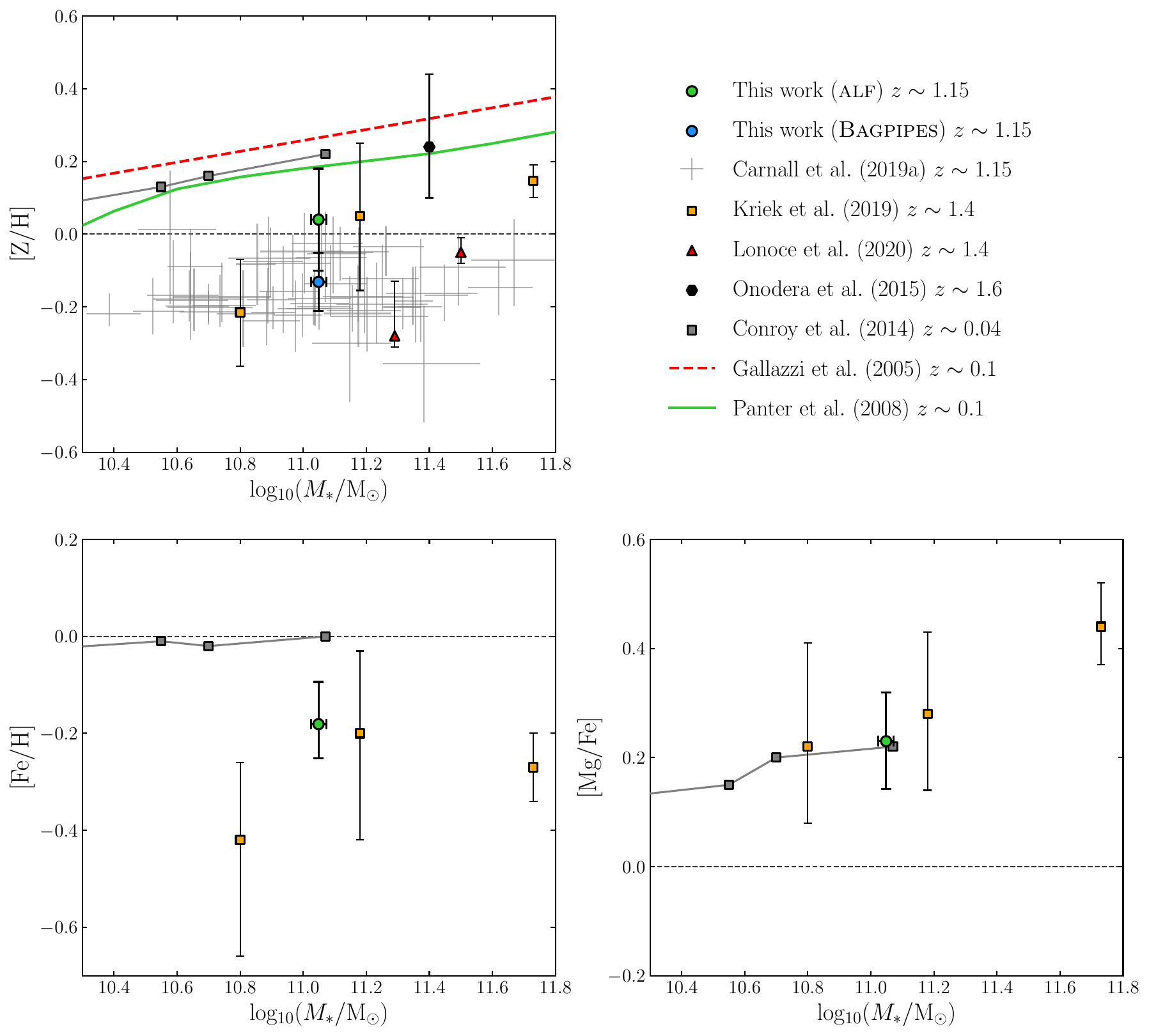}
	
    \caption{Top: stellar metallicity measurements for massive quiescent galaxies at $z\gtrsim1$, compared with the local Universe. Our results, which are broadly consistent with previous work at similar redshifts, suggest that [Z/H] was $\sim0.2-0.3$ dex lower at $z\gtrsim1$ compared with the present day. Lower left: Fe abundances for the subset of studies, including ours, that use \textsc{Alf}. We find [Fe/H] $\sim0.15$ dex lower than the local Universe. Lower right: alpha enhancement, traced by [Mg/Fe]. We find $\sim0.2$ dex enhancement, suggesting no evolution across the last $\sim8$ Gyr since $z\gtrsim1$. The dashed horizontal lines show the Solar value.}
    \label{fig:metal}
\end{figure*}

\subsection{ALF fitting of the stacked spectrum}\label{stack:fit_alf}

In addition to fitting our stacked data with \textsc{Bagpipes}, we also carried out an independent analysis with the \textsc{Alf} code\footnote{\url{https://www.github.com/cconroy20/alf}} (\citealt{Conroy2012, Conroy2018}). This decision was made, firstly to provide a cross-check on our results via a more established method, and secondly as the \cite{Bruzual2003} models assume scaled-Solar abundances. By contrast, \textsc{Alf} allows individual element abundances to vary separately, allowing us to constrain the level of alpha enhancement in our target population, a valuable indicator of their formation timescales (e.g. \citealt{Thomas2003}).

The \textsc{Alf} code is designed for fitting optical to near-IR continuum spectroscopy for old ($\gtrsim$ 1 Gyr) stellar populations. Originally designed for constraining the initial mass function, recent applications have focused on stellar metallicities, with the code having been developed to fit individual abundances for up to 19 elements. 

The code also makes use of the MILES stellar spectral library, and includes empirical stellar population models spanning $0.37{-}2.4\,\mu$m. We therefore fit our stacked spectrum across a wavelength range similar to that described in Section \ref{stack:fit_bagpipes}, this time also omitting rest-frame wavelengths from $3550{-}3700$\,\AA. We investigate the potential impact of this difference by re-running our \textsc{Bagpipes} fit whilst also excluding this wavelength range, and find this has no effect on our results.

\textsc{Alf} continuum normalises input spectra using a high-order polynomial, with one order per 100\,\AA\ of rest-frame spectral coverage, prior to fitting. The code does not currently include the capability to fit photometric data, so our stacked photometry was not used in this analysis. The code uses the MCMC sampler \textsc{Emcee} (\citealt{Foreman-Mackey2013}). We run \textsc{Alf} in simple mode, which includes 13 free parameters: redshift, velocity dispersion, stellar age (a single burst SFH is assumed), total stellar metallicity, [Z/H], and abundances for 9 individual elements, including Fe and Mg. As an additional check, we have also fitted our data using full mode in \textsc{Alf}, which results in near-identical metallicity values.

\section{Results}\label{results}

The posterior median model fitted to our stacked spectrum with \textsc{Bagpipes} is shown in red in the top panel of Fig. \ref{fig:stack}. We measure a stellar metallicity of [Z/H] = $-0.13\pm0.08$. We also measure a mean stellar age of $2.5^{+0.6}_{-0.4}$ Gyr, which, at $z=1.15$, corresponds to a mean formation time, $t_\mathrm{form} = 2.7^{+0.4}_{-0.6}$ Gyr after the Big Bang, or a formation redshift of $z_\mathrm{form} = 2.4^{+0.6}_{-0.3}$. We measure a dust curve slope consistent with \cite{Calzetti2000}, and an $A_V = 0.48\pm0.08$ magnitudes. This is in good agreement with \cite{Belli2019} and \cite{Carnall2019b} (but see appendix B of \citealt{vanderWel2021}).

With \textsc{Alf}, we obtain element abundances \hbox{[Fe/H] = $-0.18\pm0.08$} and \hbox{[Mg/H] = $0.07\pm0.09$}. The enhancement of Mg relative to Fe can be used as a proxy for alpha enhancement, and we obtain \hbox{[Mg/Fe] = $0.23\pm0.12$}. We convert from Fe and Mg abundances to total metallicity using [Z/H] = [Fe/H] + 0.94[Mg/Fe] (e.g. \citealt{Thomas2003}), for consistency with other recent work (e.g. \citealt{Kriek2019}). This yields a total metallicity of [Z/H] = $0.04\pm0.14$. Our \textsc{Bagpipes} and \textsc{Alf} measurements of [Z/H] are therefore in broad agreement, with the \textsc{Alf} posterior median value 0.17 dex higher than from \textsc{Bagpipes} (formally the two posterior distributions are in tension with $\sim1.1\sigma$ confidence). The \textsc{Alf} fitted spectrum is virtually indistinguishable from the \textsc{Bagpipes} fitted spectrum shown in Fig. \ref{fig:stack}, and is therefore not shown. Our stellar metallicity results are summarised in Fig. \ref{fig:metal}, and compared to results from the literature in Section \ref{disc:metal}.

The stellar population ages are also in broad agreement, with \textsc{Alf} returning an age of $2.4\pm0.3$ Gyr. The fact \textsc{Alf} returns a $\sim1\sigma$ higher metallicity and a slightly lower age than \textsc{Bagpipes} is consistent with our expectations, based on the different SFH models used by the two codes. When run in simple mode, \textsc{Alf} uses a single burst SFH, a simplifying assumption known to result in lower ages and higher metallicities, closer to the light-weighted values (e.g. \citealt{Conroy2013}). These issues will be further discussed in Section \ref{discussion}.

\section{Discussion}\label{discussion}

\subsection{Stellar Metallicities}\label{disc:metal}

\subsubsection{Comparisons with other results at $z\gtrsim1$}\label{disc:metalz1}

Whilst this work presents the first measurement of the average stellar metallicity of a mass-selected sample of massive quiescent galaxies at \hbox{$z>1$}, several other studies have analysed individual bright objects, or magnitude-selected samples, at similar redshifts. In Fig. \ref{fig:metal}, we compare our results for [Z/H], [Fe/H] and [Mg/Fe] with other recent work at $1.0 < z \lesssim 1.5$, as well as results from the local Universe, which will be discussed in Section \ref{disc:metalz0}.

\cite{Kriek2019} report stellar metallicities for three massive quiescent galaxies at $z\sim1.4$ via full spectral fitting of Keck-LRIS and MOSFIRE spectroscopy, also with the \textsc{Alf} code. These results are shown in \hbox{Fig. \ref{fig:metal}} with orange squares, and are in good agreement with the measurements we derive from our stacked spectrum. Their most massive object is more Mg-enhanced, leading to a higher total metallicity, consistent with the positive correlation between mass and metallicity observed in the local Universe.

\cite{Lonoce2015, Lonoce2020} also report stellar metallicities for two galaxies at $z\sim1.4$ via full spectral fitting, which are shown in the top-left panel of Fig. \ref{fig:metal} with red triangles. These results are also consistent with our findings, though it should be noted that \cite{Lonoce2020} obtain strongly contrasting results via an alternative, spectral-index-fitting analysis: [Z/H] $\sim-0.7$ and $+0.6$ for their lower and higher-mass galaxies, respectively.

\cite{Onodera2015} report a stellar metallicity of \hbox{[Z/H] = $0.24^{+0.20}_{-0.14}$} via spectral index fitting, for a stacked spectrum constructed from a magnitude-selected sample ($K < 21.5$) of 24 quiescent galaxies at $1.2 \lesssim z \lesssim 2.0$. This is considerably higher than our result at $1.0 < z < 1.3$, however, their relatively large reported uncertainties mean the two results are not strongly in tension. It should also be noted that the average mass of galaxies in the sample of \cite{Onodera2015} is 0.35 dex higher than our sample.

Finally, we show, with gray error bars, the stellar metallicities we obtained for 53 individual massive quiescent galaxies in \cite{Carnall2019b}, via \textsc{Bagpipes} full spectral fitting of intermediate-SNR ($\sim10$\,\AA$^{-1}$) VANDELS rest-frame near-UV spectroscopy. The \cite{Carnall2019b} results are in very good agreement with the new analysis we present in this work, which differs from our previous work in three key respects: the use of stacking, the inclusion of additional rest-frame optical KMOS data, and the exclusion of wavelengths, $\lambda < 3550$\,\AA, which were fitted with theoretical stellar templates in \cite{Carnall2019b}. Our new result validates the metallicities measured by our previous analysis, and hence the ages/SFHs that were the main result of that work. 

The individual-object results of \cite{Carnall2019b} shown in \hbox{Fig. \ref{fig:metal}} do not suggest a strong stellar mass-metallicity relation at $z\gtrsim1$ above log$(M_*/\mathrm{M_\odot}) = 10.5$, and are consistent with a flat relationship. However, a relatively weak mass-metallicity relation, as seen in the local Universe, is by no means ruled out. We have investigated the possibility of splitting the mass-complete sample in this work into two mass bins to further probe the mass-metallicity relation at $z\gtrsim1$, however our KMOS dataset does not provide sufficient SNR to obtain meaningful constraints when splitting our sample into two or more bins.

We conclude that our new results are generally consistent with previous work at $z\gtrsim1$. In particular, multiple full spectral fitting analyses agree on average stellar metallicities, [Z/H] $\sim-0.1$ for massive quiescent galaxies with log$_{10}(M_*/\mathrm{M}_\odot) \sim 11$ at $z\gtrsim1$. Our measurement of [Fe/H] $\sim-0.2$ is in good agreement with \cite{Kriek2019}, as is our measurement of $\sim0.2$ dex of Mg enhancement over Fe. \cite{Kriek2016, Kriek2019} show that this level of alpha enhancement suggests formation timescales of $0.2{-}1.0$ Gyr, again in good agreement with the SFHs reported by \cite{Carnall2019b}.

\subsubsection{Comparisons with the Local Universe}\label{disc:metalz0}

In the top-left panel of Fig. \ref{fig:metal}, we also show three determinations of the stellar mass-metallicity relation in the local Universe ($z \lesssim 0.1$). These three sets of results are all in reasonable agreement, however there are key differences in methodology and sample selection that should be considered when making comparisons. All three studies use data from the Sloan Digital Sky Survey (SDSS; \citealt{York2000}).

\cite{Gallazzi2005} and \cite{Panter2008} report mean stellar metallicity as a function of stellar mass for all galaxies, including both star-forming and quiescent objects. However, the dominance of quiescent galaxies at high masses in the local Universe means these results are still reasonably comparable with the $z\gtrsim1$ results shown. In addition, \cite{Gallazzi2005} report $r-$band light-weighted metallicities, rather than the mass-weighted quantities reported by the other studies. This may explain the higher metallicities they report, as their results are weighted more heavily towards younger stars.

It should also be noted however that even among studies reporting mass-weighted quantities, the SFH model adopted has the potential to influence the stellar metallicities obtained (e.g. \citealt{Carnall2019a}; \citealt{Leja2019a}). This is of particular note for \cite{Conroy2014}, who adopt a simplified two-burst SFH model that would be expected to return metallicities closer to the light-weighted values. This may explain the slightly higher metallicities found by \cite{Conroy2014} with respect to \cite{Panter2008}.

The results of \cite{Conroy2014} are the most comparable with our results at $z\gtrsim1$, both in terms of methodology and sample selection. They select quiescent galaxies by requiring no H$\alpha$ or [O\,\textsc{ii}] emission, and require that objects lie on the Fundamental Plane, using the central slice defined by \cite{Graves2010}. At \hbox{log$(M_*/\mathrm{M_\odot})\sim11$}, they report [Z/H] = 0.21. Our result therefore implies, at fixed stellar mass, $\sim0.2-0.3$ dex evolution in the average [Z/H] for massive quiescent galaxies across the $\sim8$ Gyr from $z\sim1.15$ to the present. From the lower-right panel of Fig. \ref{fig:metal}, it can be seen that, surprisingly, this 0.2-0.3 dex evolution in [Z/H] is not accompanied by any change in alpha (Mg) enhancement, with [Mg/Fe] remaining at $\sim0.2$.

Alpha elements are primarily produced by core-collapse supernovae (CCSNe), which are the end-point in the evolution of massive stars. Enrichment with alpha elements therefore rapidly follows the onset of star formation, within a few Myr. By contrast, Fe-peak elements are produced by both CCSNe and Type Ia supernovae (SNIa), which occur on  much longer ($\sim$ Gyr) timescales. This means that both Fe abundance and alpha enhancement in stars are linked to the formation timescale of the galaxy, with lower Fe, more highly alpha enhanced stellar populations expected in galaxies that formed on short timescales (e.g. \citealt{Weinberg2017}; \citealt{Kobayashi2020}).

As the number density of massive quiescent galaxies steadily rises over cosmic time, those that exist at $z\gtrsim1$ are expected to be a biased sub-sample of the local quiescent galaxy population: the earliest and fastest formed. This is often referred to as progenitor bias. We would therefore expect to see, on average, both lower Fe abundance and greater alpha enhancement for $z\gtrsim1$ quiescent galaxies, compared with the local population.

\cite{Beverage2021} have recently reported similar evolution of the quiescent galaxy population from the local Universe to $z\sim0.7$: falling [Z/H] and [Fe/H], but constant [Mg/Fe] with increasing redshift (see also \citealt{Choi2014, Leethochawalit2019}). They suggest the expected increase in [Mg/Fe] might begin to become apparent at higher redshifts, pointing to several individual-object studies that find very high [Mg/Fe] for individual ultra-massive galaxies at $z\sim2$ (\citealt{Kriek2016}; \citealt{Jafariyazani2020}). 

Our results demonstrate the expected increase in [Mg/Fe] is still not apparent at $z\gtrsim1$ for typical log$(M_*/\mathrm{M_\odot})\sim11$ quiescent galaxies. This is in agreement with \cite{Kriek2019} at $z\sim1.4$ (bottom-right panel of Fig. \ref{fig:metal}), who report higher alpha enhancement than the local Universe for only a single, extremely massive galaxy  with log$(M_*/\mathrm{M_\odot})\sim11.7$. Our results suggest the very high [Mg/Fe] values (implying ultra-short, \hbox{$\sim100$ Myr}, formation timescales) recently reported by \cite{Kriek2016, Kriek2019} and \cite{Jafariyazani2020} for ultra-massive galaxies at $z\sim1-2$ are more closely associated with their exceptionally high masses, in accordance with the well-known downsizing trend, rather than their high redshifts.

The expected change in [Mg/Fe] across the $\sim0.15$ dex interval between our [Fe/H] result and those of \cite{Conroy2014} can be estimated from fig. 12 of \cite{Kobayashi2020} to be $\sim0.1$ dex. This is of the same order of magnitude as the uncertainty we measure on our [Mg/Fe] value. This means that, whilst we do not observe any change in [Mg/Fe] between the local Universe and our results at $z\gtrsim1$, the level of [Mg/Fe] evolution predicted by \cite{Kobayashi2020} is not strongly excluded by our measurement.

A detailed understanding of these issues will require a dataset of similar quality to SDSS at $z\gtrsim1$. Several planned surveys with upcoming instruments, such as the Multi-Object Optical and Near-infrared Spectrograph (MOONS) GTO survey MOONRISE (\citealt{Cirasuolo2020}; \citealt{Maiolino2020}), will provide much larger numbers of high-SNR spectra at $z\gtrsim1$. This will allow individual metallicity determinations for large numbers of bright targets, and high-SNR stacking experiments to be performed with fine resolution in stellar mass and other key parameters, such as physical size.

\subsection{Stellar Ages}\label{disc:ages}

As discussed in Section \ref{disc:metalz1}, the stellar metallicity we derive for our stacked spectrum is in good agreement with those we obtained for individual galaxies in \cite{Carnall2019b}. In that work, we derived the following relationship between formation time, $t_\mathrm{form}$ (which we define as the time corresponding to the mean stellar age), and stellar mass for quiescent galaxies at $z\gtrsim1$:
\begin{equation}\label{eqn:tform}
\bigg(\dfrac{t_\mathrm{form}}{\mathrm{Gyr}}\bigg)\ =\ 2.56^{+0.12}_{-0.10}\ -\ 1.48^{+0.34}_{-0.39}\ \mathrm{log}_{10}\bigg(\dfrac{M_*}{10^{11}\mathrm{M_\odot}}\bigg).
\end{equation}

\noindent At the median stellar mass of \hbox{$\mathrm{log_{10}(}M_*/\mathrm{M_\odot)} = 11.05$} for the mass-complete sample from which we construct our stack, this relationship predicts a formation time of 2.5$\pm$0.1 Gyr after the Big Bang. We derive consistent results from both the \textsc{Bagpipes} and \textsc{Alf} analyses of our stack: $t_\mathrm{form} = 2.7^{+0.4}_{-0.6}$ Gyr and 2.4$\pm$0.3 Gyr respectively.

Several other recent studies have reported ages for massive quiescent galaxies at $z\gtrsim1$ using similar methodologies (\citealt{Belli2019, Estrada-Carpenter2020, Tacchella2021}). These results are summarised in fig. 10 of \cite{Tacchella2021}, demonstrating that the formation times they derive for their $z\sim0.8$ sample are earlier on average than those we derive at $z\gtrsim1$. This is the opposite to the effect that would be expected due to progenitor bias.

Due to the strong age-metallicity degeneracy in galaxy spectral shapes, discrepant metallicities is an obvious potential cause for the unexpected differences in derived ages between these studies. However, as can be seen from fig. 19 of \cite{Tacchella2021}, the average metallicity they derive is just below Solar\footnote{The average metallicity derived for the UVJ-quiescent sample in \cite{Tacchella2021} is  [Z/H] = $-0.13\pm0.02$ (private communication).}, as would be expected considering the comparison between local Universe and \hbox{$z\gtrsim1$} metallicities shown in the top panel of \hbox{Fig. \ref{fig:metal}}. The metallicities we derive in this work and \cite{Carnall2019b} are therefore fully consistent with \cite{Tacchella2021}, and are ruled out as the cause of the differences in derived ages.

The remaining likely cause for these differences lies in the SFH models used by the respective studies. In this work and \cite{Carnall2019b}, we have used the double-power-law SFH model introduced by \cite{Carnall2018}. This is a parametric model, and similar to the models used by \cite{Belli2019}, who report results consistent with \cite{Carnall2019b}. By contrast, \cite{Tacchella2021} use the non-parametric, continuity SFH prior introduced by \cite{Leja2019a}, which is known to return older stellar populations than more traditional parametric models \citep{Leja2019b}. Interestingly, \cite{Estrada-Carpenter2020} also use the non-parametric continuity prior, finding results consistent with \cite{Tacchella2021}.

It seems likely, therefore, that the underlying cause for the discrepant results in the recent literature for the formation redshifts of massive quiescent galaxies at $z\gtrsim1$ is the use of different SFH models, rather than conflicting derived metallicities. These SFH models constitute a set of prior beliefs about when and how galaxies form, introducing an unavoidable subjectivity into results obtained in the absence of strongly constraining data (e.g. \citealt{Ocvirk2006}; \citealt{Carnall2019a}; \citealt{Johnson2021}).

To evaluate the success of these subjective prior choices, one should perform cross-validation checks, testing whether results obtained via spectral fitting are consistent with other, better established results in the literature (e.g. \citealt{Wuyts2011, Carnall2019a}). In this case, we would wish to use the SFHs derived for $z\gtrsim1$ massive quiescent galaxies to predict the stellar-mass function for quiescent galaxies at $z\sim2-4$, then to compare these predictions with direct observations of $z>2$ mass functions and/or number densities.

To perform this challenging experiment, we are currently missing two vital components. Firstly, the number of massive quiescent galaxies at $z\gtrsim1$ with high-SNR rest-frame UV-optical spectroscopy is currently no more than a few hundred. This, in combination with the significant uncertainties on individual recovered SFHs, leads to huge statistical uncertainties in predicted mass functions at $z>2$. To make useful predictions, vastly larger samples will be required, numbering tens to hundreds of thousands of galaxies. A number of upcoming surveys, such as MOONRISE (\citealt{Maiolino2020}), will provide the necessary datasets over the next $\sim5$ years.

The second missing component is a precise determination of the observed stellar-mass function for massive quiescent galaxies at \hbox{$z>2$}, with current studies suffering from considerable uncertainties in sample selection (e.g. \citealt{Schreiber2018, Carnall2020}). Large imaging surveys with JWST (e.g. PRIMER; \citealt{Dunlop2021}) will provide the extremely deep, high-resolution infra-red imaging necessary to select mass-complete samples of quiescent galaxies with confidence out to the highest redshifts, resulting in precise constraints on quiescent galaxy stellar-mass functions at $z>2$.

Once these datasets are in place, it should be possible to clearly distinguish between the parametric and non-parametric SFH models that currently return substantially different formation redshifts for massive quiescent galaxies at $z\gtrsim1$.

\section{Conclusion}\label{conclusion}

In this work, we have combined data from the VANDELS survey with new KMOS observations to construct a representative stacked spectrum for quiescent galaxies at $1.0 < z < 1.3$ with $\mathrm{log}_{10}(M_*\ /\ \mathrm{M_\odot}) > 10.8$, covering rest-frame wavelengths, \hbox{$\lambda=2500{-}6400$\,\AA.} The stacked spectrum is shown in Fig. \ref{fig:stack}. We also report 25 new spectroscopic redshifts (Table \ref{table:redshifts}) from our KMOS data.

We fit our stacked spectrum with \textsc{Bagpipes}, obtaining a stellar metallicity, [Z/H] = $-0.13\pm0.08$. We also obtain a formation time, $t_\mathrm{form} = 2.7^{+0.4}_{-0.6}$ Gyr after the Big Bang, corresponding to a formation redshift of $z_\mathrm{form}=2.4^{+0.6}_{-0.3}$. Both of these results are consistent with the results we presented in \cite{Carnall2019b}, which were obtained by fitting only VANDELS data, and working with individual spectra rather than stacking. We also fit our stacked spectrum with the \textsc{Alf} code, obtaining a consistent result for [Z/H], as well as an iron abundance, [Fe/H] = $-0.18\pm0.08$, and alpha enhancement, [Mg/Fe] = $0.23\pm0.12$ (see Fig. \ref{fig:metal}).

By comparing our results at $z\gtrsim1$ with results from the local Universe ($z\lesssim0.1$), we demonstrate that the average [Z/H] for \hbox{$\mathrm{log}_{10}(M_*\ /\ \mathrm{M_\odot}) \sim 11$} quiescent galaxies has risen by $\sim0.3$ dex across the $\sim8$ Gyr since $z\gtrsim1$, whereas [Fe/H] has risen by $\sim0.15$ dex. However, the alpha enhancement, [Mg/Fe], we measure is the same as found by \cite{Conroy2014} in the local Universe, implying no evolution at fixed stellar mass across at least the last $\sim8$ Gyr.

Given that $z\gtrsim1$ massive quiescent galaxies are a biased sub-sample of the local quiescent population (those that formed fastest and earliest), we would expect them to have lower [Fe/H], as observed, but we would also expect greater alpha enhancement. Our finding of no redshift evolution in [Mg/Fe] at \hbox{$\mathrm{log}_{10}(M_*\ /\ \mathrm{M_\odot}) \sim 11$} from $z\gtrsim1$ to the present is in agreement with the results of \cite{Kriek2019} at $z\sim1.4$ and \cite{Beverage2021} at $z\sim0.7$.

This result suggests that the highly alpha enhanced, ultra-massive ($\mathrm{log}_{10}(M_*\ /\ \mathrm{M_\odot}) \gtrsim 11.5$) galaxies recently reported at $z\sim1-2$ by some authors (e.g. \citealt{Kriek2016, Kriek2019, Leethochawalit2019}) are highly alpha enhanced due to their extreme masses, in accordance with the well-known downsizing trend, rather than being typical of the $z\gtrsim1$ quiescent population.

The model of \cite{Kobayashi2020} predicts a relatively modest change in [Mg/Fe] of $\sim0.1$ dex across the $\sim0.15$ dex [Fe/H] interval separating our results from the local-Universe measurements of \cite{Conroy2014}. This level of [Mg/Fe] evolution is not strongly ruled out by our measurement, meaning that stronger constraints from higher-quality data at $z\gtrsim1$ will be necessary to confidently determine whether the level of [Mg/Fe] evolution observed between the local Universe and $z\gtrsim1$ is in agreement with theoretical predictions.

Recently, \cite{Tacchella2021} have highlighted differences in the average formation redshifts measured for $z\gtrsim1$ massive quiescent galaxies by several recent studies. In particular, they, in agreement with \cite{Estrada-Carpenter2020}, find earlier formation than \cite{Belli2019} and \cite{Carnall2019b}. Given the latter two studies are at higher redshifts, this is the opposite to the effect that would be expected due to progenitor bias. 

We demonstrate that the metallicities recovered by \cite{Tacchella2021} at $z\sim0.8$ are consistent with the results of this work and \cite{Carnall2019b}. This means discrepancies in recovered metallicities are ruled out as the cause of the differences in recovered ages. We therefore conclude that the differences in formation redshifts obtained by different authors are likely due to the use of different star-formation-history models, with \cite{Estrada-Carpenter2020} and \cite{Tacchella2021} using the \cite{Leja2019a} continuity non-parametric SFH model, and \cite{Belli2019} and \cite{Carnall2019b} using more-traditional parametric models.

To determine which of these models more accurately represents the SFHs of $z\gtrsim1$ massive quiescent galaxies, we will require much larger statistical samples at \hbox{$z\gtrsim1$} with high-SNR rest-frame UV-optical spectroscopy, as will be provided by upcoming instruments such as MOONS \citep{Cirasuolo2020}. This will allow us to make firm predictions for $z>2$ quiescent-galaxy stellar-mass functions via spectral fitting with different SFH models, which can be compared with the precise $z>2$ stellar-mass functions that will be provided by data from the upcoming \textit{James Webb Space Telescope}.

\section*{Acknowledgements}

The authors would like to thank Corentin Schreiber, David Maltby and Omar Almaini for valuable advice about KMOS observing strategies and SNRs. We would also like to thank Michael Hilker at the ESO helpdesk for assistance with KMOS data reduction. The authors would also like to thank the anonymous referee for their helpful comments. A. C. Carnall would like to thank the Leverhulme Trust for their support via the Leverhulme Early Career Fellowship scheme. A. Cimatti acknowledges support from the grant PRIN MIUR 2017$-$20173ML3WW\_001. B. Garilli acknowledges support from the grants Premiale MITIC 2017 and INAF PRIN "Mainstream 2019". Based on observations made with ESO Telescopes at the La Silla or Paranal Observatories under programme ID(s) 194.A-2003(A-T) and 0104.B-0885(A).

\section*{Data Availability}

The VANDELS survey is a European Southern Observatory Public Spectroscopic Survey. The full spectroscopic dataset, together with photometric catalogues and derived quantities are available from \url{http://vandels.inaf.it}, as well as from the ESO archive \url{https://www.eso.org/qi}. The KMOS data used in this work is available from the ESO archive under programme ID 0104.B-0885(A). Reduced KMOS data products, our stacked spectrum and model posteriors may be made available upon reasonable request.

\bibliography{carnall2021}
\bibliographystyle{aasjournal}

\end{document}